\documentclass[final,1p,times]{elsarticle}

\usepackage{graphicx}
\usepackage{amssymb}
\usepackage{amsthm}
\usepackage{lineno}
\usepackage{float}

\journal{Nuclear Physics A}

\begin{document}

\begin{frontmatter}

\title{Measurement of the ratio of the sixth order to the second order cumulant of net-proton multiplicity  distributions in relativistic heavy-ion collisions}

\author{Lizhu Chen$^{1, 2}$, Zhiming Li $^2$, Fenping Cui$^1$ and  Yuanfang Wu$^2$}
\address{$^1$ School of Physics and Optoelectronic Engineering, Nanjing University of Information Science and Technology, Nanjing 210044, China}
\address{$^2$ Key Laboratory of Quark and Lepton Physics (MOE) and Institute of Particle Physics, Central China Normal University, Wuhan 430079, China}

\begin{abstract}
We investigate the measurement of the sixth order cumulant and its ratio to the second order cumulant ($C_6/C_2$) in relativistic heavy-ion collisions. The influence of statistics and different  methods of centrality bin width correction on $C_6/C_2$ of net-proton multiplicity distributions is demonstrated. There is no  satisfactory method to extract  $C_6/C_2$ with the current statistics recorded at lower energies by STAR at RHIC. With statistics comparable to the expected statistics at the planned future RHIC  Beam Energy Scan II (BES II), no energy dependence of $C_6/C_2$  is observed in central collisions using the UrQMD model. We find if the transition signal is as
strong as predicted by the PQM model, then it is hopefully observed at the upcoming RHIC BES II.
\end{abstract}

\end{frontmatter} 


\section{Introduction}

Uncovering the structure of the QCD phase diagram is one of the major goals in studying relativistic heavy-ion collisions.
Ratios of cumulants of conserved quantities, such as net-baryon, net-charge and net-strangeness numbers,  can exhibit large fluctuations near the QCD phase transition~\cite{QCD-1, QCD-2, QCD-3, QCD-4, QCD-5, QCD-6, QCD-7, QCD-8}.
Due to the finite size effects and critical slowing down~\cite{slowing-1, slowing-2},  such significant enhanced fluctuations would be suppressed in experiment. Instead, the phenomena of oscillatory  behaviors, such as the non-monotonic and sign change, are the particularly interesting  signals which could  be directly measured in experiments~\cite{ STAR-proton, STAR-charge,phenix-charge}. Theoretically, it has been found that the sixth order to the second order  cumulant ratio ($C_6/C_2$)
undergo a significant sign change near the QCD phase transition~\cite{QCD-4, QCD-9, PQM, panxue,o4}. Consequently,
$C_6/C_2$ is a promising observable and it can be measured from RHIC to up to LHC energies~\cite{PQM,star,lhc}. Unfortunately, comparing to the other lower order cumulant measurements,
the price of measuring $C_6/C_2$ in an experiment is the larger number of required events and the stronger effects of non-critical background contributions.

Currently, there are many discussions on the effects of  the non-critical background contributions for cumulants.
 Due to the global  conservation, the cumulants of the conserved quantities are substantially suppressed~\cite{conservation}. The effects of the acceptance cuts as well as efficiency corrections drastically influence the measured cumulants~\cite{efficiency-1,efficiency-2}. The initial size fluctuation is also  one of the important non-critical effects which exists in all experiment event variables~\cite{initial-1, Xiaofeng-JPG, chenlz-JPG}.
What we want to emphasize here is that the techniques to reduce those non-critical effects should be studied carefully case by case for different cumulants.  With the STAR detector at RHIC, to suppress the effects of auto-correlation in the measurement, the centrality for the measurement of cumulants of the net-proton multiplicity distributions is determined by number of charged kaons and pions produced in the final state within $|\eta|<1.0$~\cite{STAR-proton, Xiaofeng-JPG}, while it is defined by final state multiplicity within $0.5<|\eta|<1.0$ for the measurement of cumulants  of the net-charge multiplicity distributions~\cite{STAR-charge, Nihar-charge}. 
Due to the detector setup of PHENIX at RHIC, there is no issue of auto-correlation for the measurement of cumulants  of the net-charge multiplicity distributions~\cite{phenix-charge, PHENIX-c}.
 For $C_6/C_2$, the effects of the auto-correlation and centrality resolution should be different at LHC and
RHIC ~\cite{star, lhc}. In this paper, we will focus on discussing the measurement of $C_6/C_2$  with the STAR detector at RHIC with BES I data.

 The  Centrality Bin Width Correction (CBWC) method is applied to reduce  initial size fluctuation on the measurement of the cumulants in heavy-ion collisions~\cite{CBWC-STAR}.
There are two key points when applying the CBWC method:  one is that the statistics in the selected centrality bin width should be sufficient to satisfy the Central Limit Theorem (CLT) ~\cite{CLT} and the other one is  that the selected centrality bin width should optimally reduce the initial size fluctuation.
With sufficient statistics, it has been found that the values of $C_4/C_2$ are consistent based on  the CBWC methods in each $\delta 1\%$ centrality bin width and in each of $N_{ch}$ ~\cite{chenlz-JPG}. Since the chosen centrality bin width can vary, the number of events that are required by the CLT should be dependent on the chosen centrality bin width used by CBWC method. For example,  within transverse momentum  range $0.4<p_{T}<0.8$ GeV/$c$, the required statistics for $C_4/C_2$ of
net-proton number is 15 million (15M) based on CBWC method in each of $N_{ch}$, while 1M is enough if CBWC method is applied  for each centrality bin with a width of  $\delta 1\%$~\cite{chenlz-JPG}.
 For  $C_6/C_2$,  the resolution with different centrality bin widths used by CBWC method  and the corresponding required statistics should be determined.

The statistics   is one of the most crucial issues for the analysis of $C_6/C_2$.
 Even with the anticipated statistics of the upcoming RHIC BES II, the statistical uncertainties of $C_6/C_2$ will be still large.
 Currently, $C_6/C_2$ obtained from Lattice QCD also exhibits large uncertainties. The QCD based model, Polyakovloop extended Quark Meson (PQM) model, shows a clear  change on the energy dependence of $C_6/C_2$~\cite{PQM}. We would further study whether this signal can be observed with the estimated  uncertainties of $C_6/C_2$  within the upcoming RHIC BES II.

In this paper, we will start off the discussions from the statistical uncertainty of $C_6/C_2$ measured by STAR at RHIC in mid-rapidity with two transverse momentum  intervals, $0.4<p_T<0.8$ GeV/$c$ and $0.4<p_{T}<2.0$ GeV/$c$, respectively. Results of $C_6/C_2$ with statistics lower than required by the CLT are also discussed.
 In section 3, we will demonstrate the statistics dependence of $C_6/C_2$ obtained from CBWC method with various centrality bin widths, $\delta2.5\%$, $\delta1\%$, $\delta0.2\%$ centrality,  and each of $N_{ch}$, using the UrQMD transport model at $\sqrt{s_{NN}}=$ 11.5 GeV in Au + Au collisions.
 An appropriate centrality bin width for the CBWC method is suggested.
In section 4, the effect of different  centrality determination methods, which would lead to different centrality resolutions,  for $C_6/C_2$ will be discussed. The energy dependence of $C_6/C_2$ of the net-proton multiplicity distributions using the  UrQMD transport model will  be discussed as well. Due to limited statistics in the recorded data by STAR during BES I, we will study if the theoretical predicted signal can be observed within uncertainties of $C_6/C_2$  at future upcoming RHIC BES II.
Finally, the results are summarized in section 6.

\section{Statistical uncertainty of $C_6/C_2$ with the STAR detector at RHIC}

 The large amount of required events is one of the most critical issues in the analysis of higher order cumulants. For example, no signal of an energy dependence for $C_4/C_2$ of net-charge multiplicity distributions  is observed within large uncertainties measured by STAR and PHENIX during BES I~\cite{STAR-charge,phenix-charge},  therefore,  it will be interesting to study it at the upcoming  RHIC BES II. Comparing to the fourth order cumulant measurement, the required statistics for $C_6/C_2$ is even larger. For $C_6/C_2$ , we should first of all determine the amount of events  that are required by the CLT.
 It is dangerous to do any data analysis if the available statistics is lower than  that required by CLT. Secondly, we need to estimate if the  transition signal can be observed within the  statistics from RHIC BES I and II.

\begin{figure}[htmp]
\centering
\includegraphics[width=1.8in]{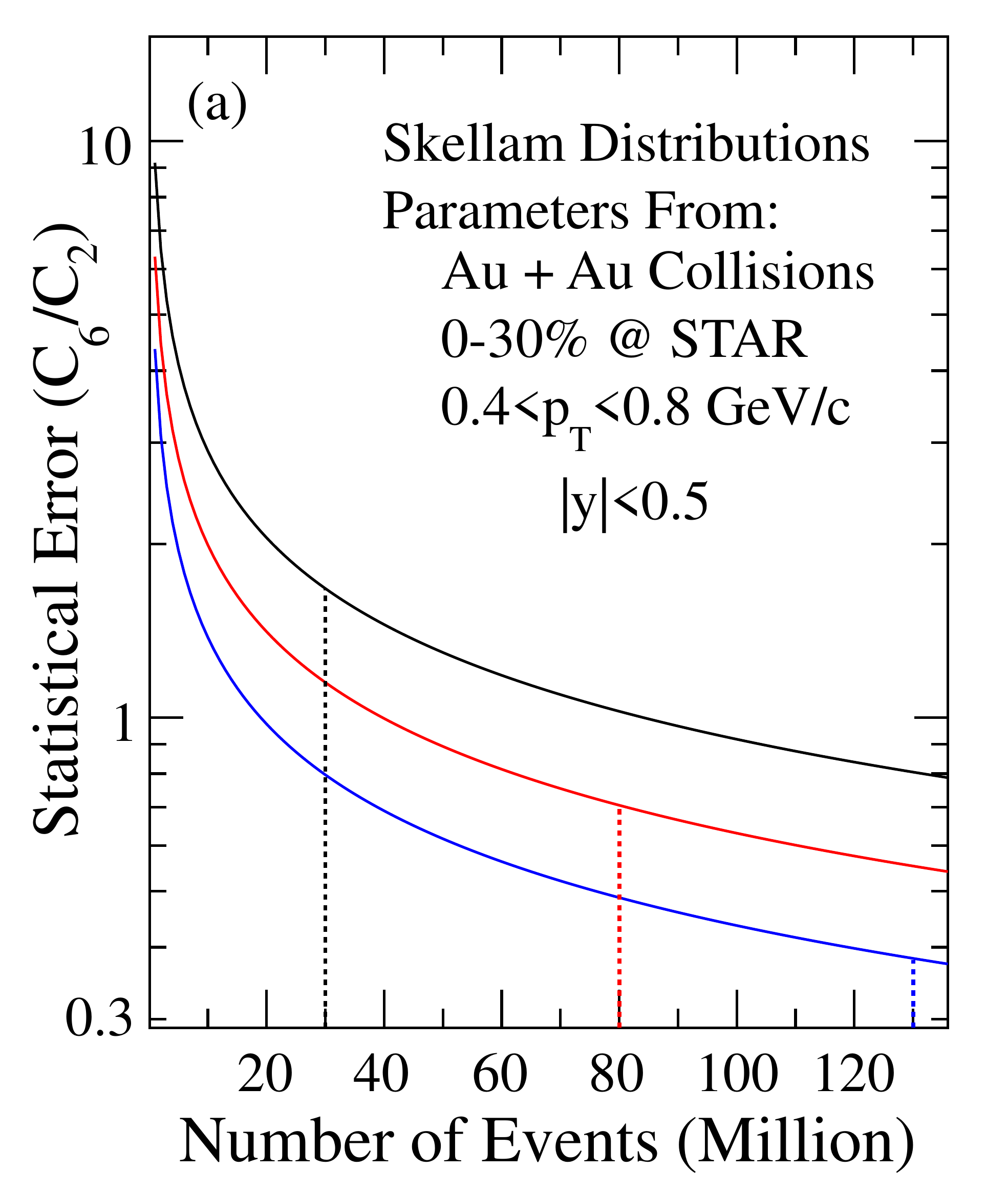}
\includegraphics[width=1.8in]{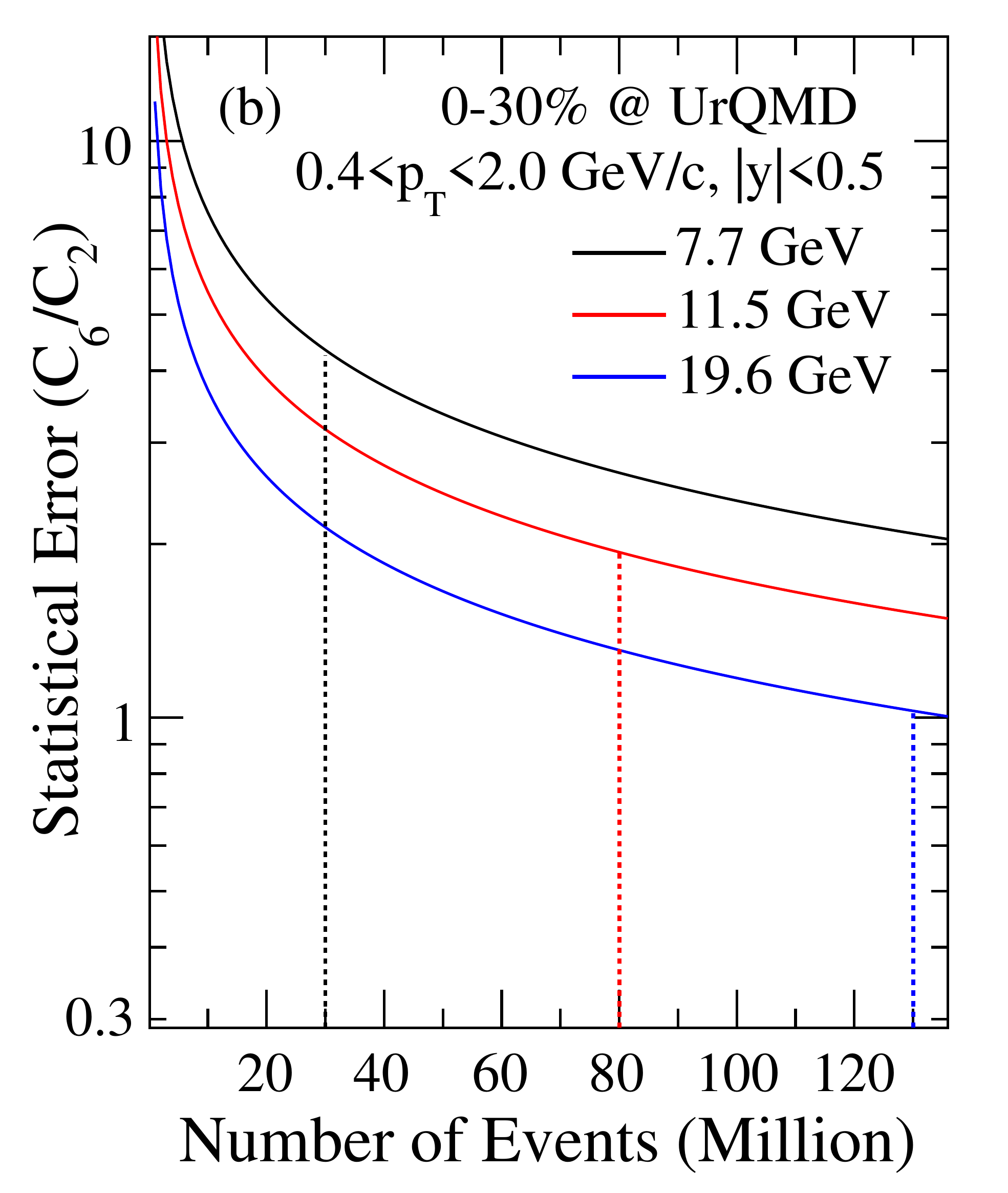}
\caption{\label{C6_error-simulation}(Color online) The  statistical uncertainties of $C_6/C_2$ with Skellam distributions calculated with the delta theorem method.  The input parameters  of (a) are taken from $\left<N_{p}\right>$ and $\left<N_{\bar{p}}\right>$ in 0-30\% central collisions in Au + Au collisions measured by STAR within $0.4<p_{T}<0.8$ GeV/$c$~\cite{STAR-proton}, and (b) are obtained using the UrQMD model within  $0.4<p_{T}<2.0$ GeV/$c$.
The number of events of vertical dashed lines are  the expected number of events at RHIC BES II.
}
\end{figure}
In order to study the  influence of statistics to the $C_6/C_2$ measurement, Fig.~\ref{C6_error-simulation} shows the statistics dependence of the statistical uncertainties of $C_6/C_2$ of net-proton multiplicity distributions using Skellam distributions~\cite{Skellam-1, Skellam-2, Skellam-3}. The statistical uncertainties are obtained using the delta theorem method~\cite{xiaofeng-delta}.  The values of two parameters using for simulations of Skellam distributions in Fig.~\ref{C6_error-simulation}(a) are 12.65,  0.12 at $\sqrt{s_{NN}}=7.7$ GeV, 9.80, 0.35 at $\sqrt{s_{NN}}=11.5$ GeV, and 7.57, 0.83 at $\sqrt{s_{NN}}=19.6$ GeV. These values are  efficiency corrected $\left<N_{p}\right>$ and  $\left<N_{\bar{p}}\right>$ in 0-30\% central collisions in Au + Au collisions measured by STAR within $0.4<p_{T}<0.8$ GeV/$c$~\cite{STAR-proton}.
 As a comparison,
the numbers of protons with the  acceptance, $0.4<p_{T}<2.0$ GeV/$c$, in right panel, are obtained using the UrQMD model~\cite{UrQMD}.  The number of events recorded in 0-30\% central collisions at the lower RHIC BES I energies are about 1.2M, 2.5M, and 6.0M at $\sqrt{s_{NN}}$=7.7, 11.5, and 19.6 GeV, respectively~\cite{STAR-proton}.
  Fig.~\ref{C6_error-simulation} shows that the statistical uncertainties are large especially with the current available statistics from BES I.  Even if the transition signal is as strong as the theoretical prediction, it can be hidden by the uncertainties.
  At $\sqrt{s_{NN}}=19.6$ GeV, the number of recorded minimum bias events by STAR is about 20M~\cite{STAR-proton}. Thus, there are about 6M events in 0-30\% central collisions. The uncertainties of $C_6/C_2$ in Fig.~\ref{C6_error-simulation}(a) with 6M events  is about 1.8 at $\sqrt{s_{NN}}=19.6$ GeV. It is already larger than value of $C_6/C_2$ obtained from Lattice QCD and the PQM model~\cite{QCD-9, PQM}.

The statistics with vertical dashed lines are the proposed numbers of events in 0-30\% central collisions to be recorded by STAR for the BES II. They are about 30M, 80M, and 130M  at 7.7, 11.5, and 19.6 GeV, respectively~\cite{BESII-statistics, BESII-STAR}.
Then the corresponding statistical uncertainties  in Fig.~\ref{C6_error-simulation}(a)  at these three energies are
1.68,  0.71,  and 0.39.  It seems the transition signal is hopeful to be observed especially at $\sqrt{s_{NN}}=19.6$ GeV.

Fig.~\ref{C6_error-simulation} shows that  the statistical uncertainties are large with current available statistics from the RHIC BES I. However, we can still study the existence of  the energy dependency based on a $3\sigma$ confidence level.  According to the CLT, the observables would take the shape of a normal distribution and the mean of the distribution is equal to the expectation value.

\begin{figure}[htmp]
\centering
\includegraphics[width=3.0in]{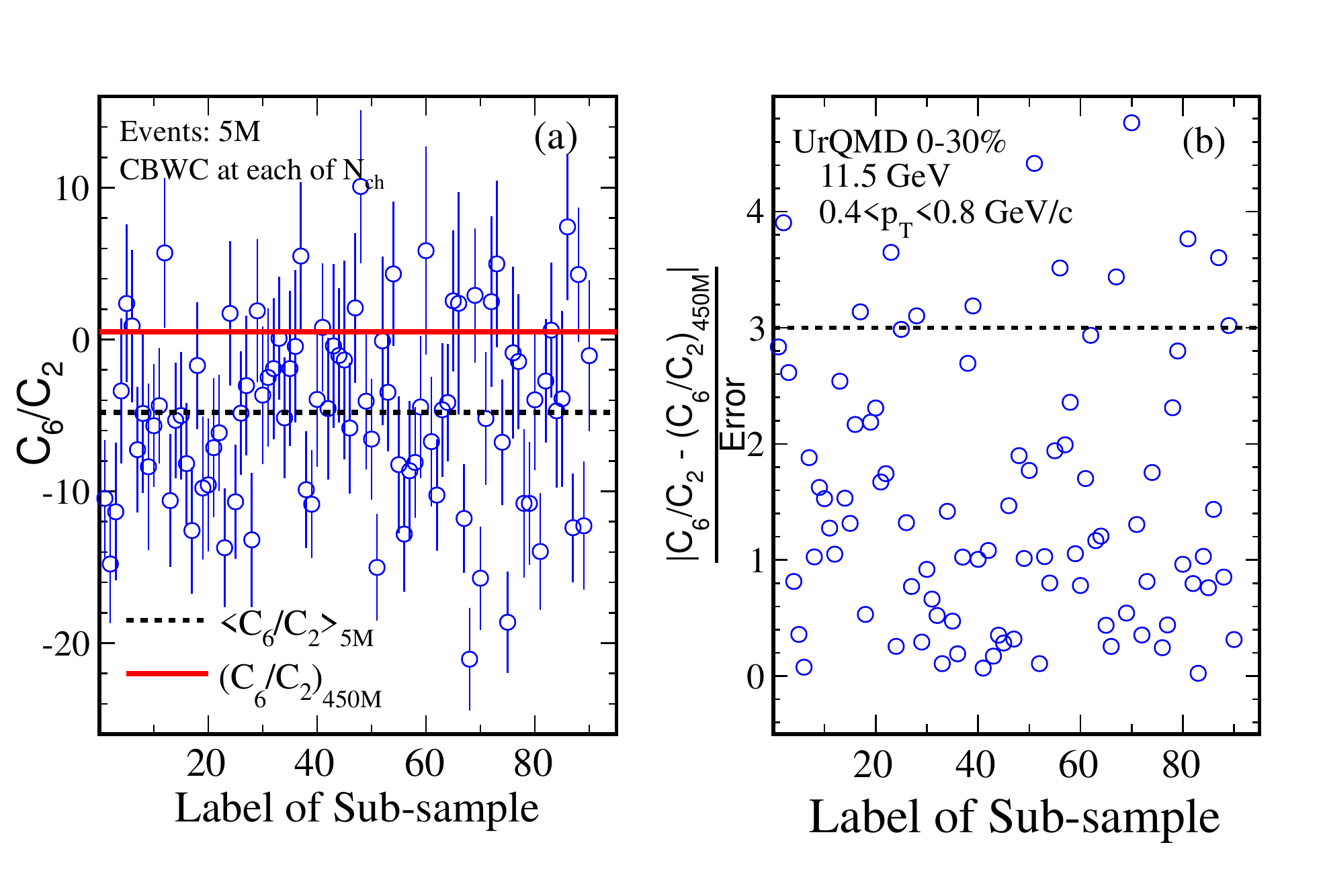}
\caption{\label{C6_sigma}(Color online) The left panel shows  $C_{6}/C_{2}$ of net-proton multiplicity distributions calculated by using the CWBC method in each of $N_{ch}$ in 0-30\% central collisions with the UrMQD model, containing 5M minimum bias events in each sub-sample. The black dashed line is the average of these 90 sub-samples, while the red solid line represents $C_{6}/C_{2}$ in 0-30\% obtained directly from the 450M minimum bias events data sample, $\left(C_6/C_2\right)_{450M}$.
The right plot demonstrates the deviation of each $C_6/C_2$ from $\left(C_6/C_2\right)_{450M}$.}
\end{figure}

We generated 450M minimum bias events at $\sqrt{s_{NN}}$  =  11.5 GeV using the UrQMD model, which is a much larger number than required by the CLT. The whole data  are randomly divided into 90 sub-samples with 5M events each, which are comparable in number of events to the recorded STAR BES I data. Fig.~\ref{C6_sigma}(a) shows $C_6/C_2$ for 0-30\% central collisions obtained from each sub-sample based on the CBWC method using each $N_{ch}$.
 The errors are obtained using the delta theorem method. It shows that $C_6/C_2$ randomly fluctuates around its mean, $\left<C_6/C_2\right>_{5M}$, which is depicted as black dashed line.
However, it also shows that  $\left<C_6/C_2\right>_{5M}$ is appreciably lower than $\left(C_6/C_2\right)_{450M}$ indicated by the red solid line, which is $C_6/C_2$ for 0-30\% central collisions obtained directly from the whole data sample (450 M). Thus, $C_6/C_2$ is systematically under-estimated in case of insufficient statistics.

To quantify the deviation of measured $C_6/C_2$  to  $\left(C_6/C_2\right)_{450M}$, $\left|C_6/C_2-\left(C_6/C_2\right)_{450M}\right|/\sigma_{C_6/C_2}$ is shown in Fig.~\ref{C6_sigma}(b).
 About 13\% of the data points are larger than 3.0, while in case of sufficient statistics only around 0.3\% of the results would be larger than 3.0. It is dangerous to discuss any physics results if the statistics is smaller than required by the  CLT.

In order not to under estimate the results, on one hand a larger amount of events could be collected by the experiments. On the other hand,
  $C_6/C_2$ shown in Fig.~\ref{C6_sigma}(a) is  obtained based on the CBWC method using each of $N_{ch}$, in which the amount of events  is much smaller than the entire data sample.
   Therefore, if a wider centrality bin width, which has the similar resolution as each $N_{ch}$, is chosen to extract $C_6/C_2$, the effect of under-estimation can be reduced.
   Given the available and expected number of events for RHIC BES I and BES II, the  CBWC method with an appropriate centrality bin width for the analysis of $C_6/C_2$ is discussed in the following section.


\section{$C_{6}/C_{2}$ obtained by using the  CBWC method with different centrality bin widths}

\begin{figure*}[htmp]
\includegraphics[width=5.0in]{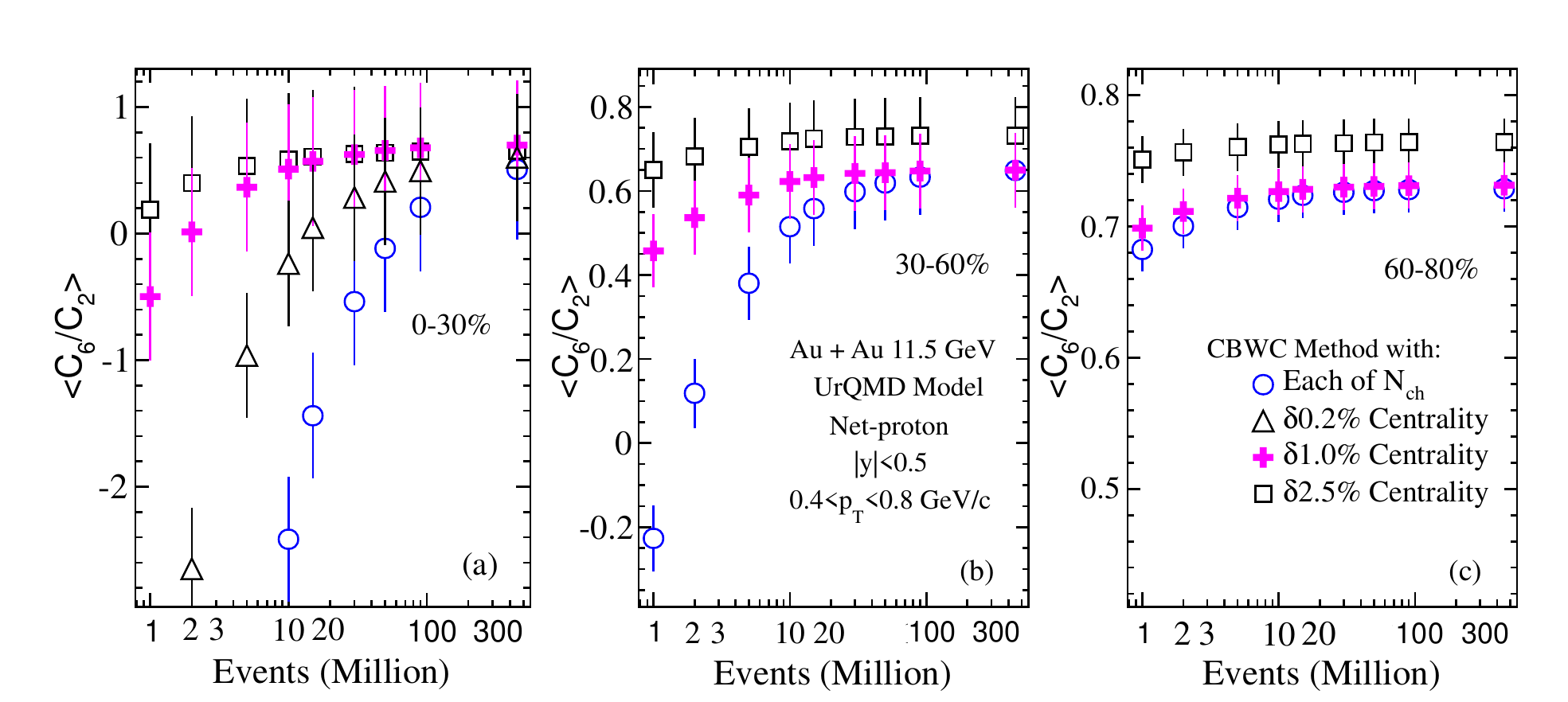}
\caption{ \label{C6C2_CBWC_centrality}  (Color online) Statistical dependence of $\left<C_{6}/C_{2}\right>$ of the net-proton multiplicity distributions obtained by using the CBWC method with various centrality bin widths:  each of $N_{ch}$ (blue open circles), $\delta$0.2\% (black open triangles), $\delta$1\% (pink solid crosses), and $\delta2.5\%$ (black open squares),
in  $0-30\%$, $30-60\%$, and $60-80\%$ centrality, respectively.  The data are generated by using the UrQMD model in Au + Au collisions at $\sqrt{s_{NN}}=$11.5 GeV.}
\end{figure*}

The data generated by  the UrQMD model at $\sqrt{s_{NN}}$ = 11.5 GeV with 450M minimum bias events are used to study $C_6/C_2$ obtained from various centrality bin widths of the CBWC method.
Similar with the previously published study on  $\kappa\sigma^2$ ($C_4/C_2$)~\cite{chenlz-JPG}, we randomly divide the total number of events into 450, 225, 90, 45, 30, 15, 9, and 5 sub-samples.  Correspondingly, the number of events
 for each kind of sub-sample are 1M, 2M, 5M, 10M, 15M, 30M, 50M, and 90M, respectively.  For each given sub-sample, we only divide it into 3 typical centralities, central collisions (0-30\%), mid-central collisions (30-60\%) and peripheral collisions (60-80\%), due to the large statistical uncertainty of $C_6/C_2$.

 Fig.~\ref{C6C2_CBWC_centrality} shows the statistical  dependence of $\left<C_6/C_2\right>$ of the net-proton multiplicity distributions  obtained by using the CBWC method  with different centrality bin widths: each of $N_{ch}$ (blue open circles)), $\delta$0.2\% (black open triangles), $\delta$1\% (pink solid crosses), and $\delta2.5\%$ (black open squares), respectively.
 The uncertainties of $(C_6/C_2)_i$ in each sub-sample are  obtained using the delta theorem method. Then the uncertainties of $\left<C_6/C_2\right>$ are obtained from  error propagation. The uncertainties of each point are very close to each other since the simulated data used for all of the data points are identical except that they are divided into different numbers of sub-samples.

  Fig.~\ref{C6C2_CBWC_centrality}(a) shows  that $\left<C_6/C_2\right>$ significantly increases as the total number of events increases applying the CBWC method in each of $N_{ch}$, or in each $\delta$0.2\% centrality bin.  With 90M events, $\left<C_6/C_2\right>_{90M}$ is still systematically lower than $\left(C_6/C_2\right)_{450M}$, which means that we can not obtain a reliable $C_6/C_2$ below 90M events (nearly 100M)  in 0-30\% central collisions.
 Consequently, it is dangerous to apply the CBWC method in each of $N_{ch}$ bin, or in each $\delta$0.2\% centrality bin, for $C_6/C_2$ if the statistics is lower than 100M events at low RHIC  energies. Otherwise,  $C_6/C_2$ in central collisions is systematically under-estimated.

 The pink solid crosses and black open squares in Fig.~\ref{C6C2_CBWC_centrality}(a) show $\left<C_6/C_2\right>$ as a function of number of events based on the CBWC method for each $\delta1\%$ and $\delta2.5\%$ centrality bin width, respectively. Below 10M minimum bias events, $\left<C_6/C_2\right>$   still increases as the number of events increases.
 Above 10M minimum bias events, the statistical dependence of $\left<C_6/C_2\right>$ is small based on CBWC method for
 each $\delta1\%$ and $\delta2.5\%$ centrality bin width. Furthermore,   above 10M minimum bias events, Fig.~\ref{C6C2_CBWC_centrality}(a) also shows that values of $C_6/C_2$ based on these two methods are both close to that obtained from each  of $N_{ch}$ with 450M minimum bias events. The difference of suppressing the initial size fluctuation is not clear due to large uncertainties. It
   can be further examined in mid-central and peripheral collisions shown in Fig.~\ref{C6C2_CBWC_centrality}(b) and~\ref{C6C2_CBWC_centrality}(c).

Above 10M events, Fig.~\ref{C6C2_CBWC_centrality}(b) and~\ref{C6C2_CBWC_centrality}(c) show that
$\left<C_6/C_2\right>$ obtained from the $\delta1\%$ centrality bin is close to $\left(C_6/C_2\right)_{450M}$. However,  $\left<C_6/C_2\right>$ obtained from $\delta2.5\%$ centrality bin is systematically larger than that obtained from the other two cases. It implies that  the effectiveness of reducing the initial size fluctuation is equal for the CBWC method in each of $N_{ch}$ and each $\delta1\%$ centrality bin, while the remnant  of initial size fluctuations based on $\delta2.5\%$ centrality bin is larger than that from each of the  $N_{ch}$ bin. Consequently, the CBWC method in each $\delta1\%$ centrality bin is a better choice for $C_6/C_2$ analysis if the number of minimum bias events is above 10M.

Let's summarize the behavior of  $C_6/C_2$ obtained from different centrality bin widths.
The effects of reducing the initial size fluctuations are similar for the CBWC method in each of the $N_{ch}$ bin and each $\delta1\%$ centrality bin. The  CBWC method in each $\delta2.5\%$ or even larger centrality bin  are  not appropriate.
450M events are of course enough for the CBWC method at each of the $N_{ch}$ bin. With statistics of 450M events, values of  $C_6/C_2$ are close to each other  in all of  three typical centralities, based on the CBWC method in each of the  $N_{ch}$ bin and each $\delta1.0\%$ centrality bin. However,
if $C_6/C_2$ is obtained from each of the $N_{ch}$ bin, 90M events are not enough. It indicates that the CBWC method in each of the $N_{ch}$ bin can not be applied in the analysis of the available BES I data and the future expected BES II data.
 If the $\delta1\%$ centrality bin is applied for $C_6/C_2$, the required number of events should be at least 10M minimum bias events. Therefore,  it is not suitable to study $C_6/C_2$ of the net-proton multiplicity distributions at $\sqrt{s_{NN}} = $ 7.7 and 11.5 GeV using STAR BES I data.   With the expected number of events in BES II, we can  obtain a reliable measurement of $C_6/C_2$ based on the CBWC method in each $\delta1\%$ centrality bin.

\section{Energy dependence of $C_6/C_2$ in UrQMD model}
Before discussing the energy dependence of $C_6/C_2$, we want to recall  the effects of different centrality determination methods on the measurement of cumulants of the net-proton multiplicity distribution, which has been studied for the measurement of  $C_4/C_2$~\cite{Xiaofeng-JPG}. It shows that the variation of different centrality determination methods, which can lead to different centrality resolutions, could also  affect the measured cumulants.  The more produced  particles in the final state that  are used to determine the centrality, the
better the  centrality resolution and the smaller fluctuations of the initial geometry we can obtain. For  $\kappa\sigma^2$ and  $S\sigma$ of the net-proton multiplicity distributions,
\cite{Xiaofeng-JPG} shows that $\kappa\sigma^2$ is more sensitive to the centrality resolution  than $S\sigma$, and it has a larger effect in more peripheral collisions as well as  at lower energies.  The even higher order of $C_6/C_2$ should also be dependent on the centrality determination.

\begin{table}[htmp]
\centering
\caption{Number of minimum bias events generated using the UrQMD model for Au + Au collisions.
\label{UrQMD-statistics}}
\begin {tabular}{|c|c|c|c|c|c|c|c|} \hline
Energy (GeV)  & 7.7  & 11.5 & 19.6 & 27 & 39 & 62.4 & 200\\
\hline
Number of Events (Million)  & 136  & 450 & 107 & 83 & 137 & 37 & 57\\
\hline
\end{tabular}
\end{table}

\begin{figure}[htmp]
\centering
\includegraphics[width=5.5in]{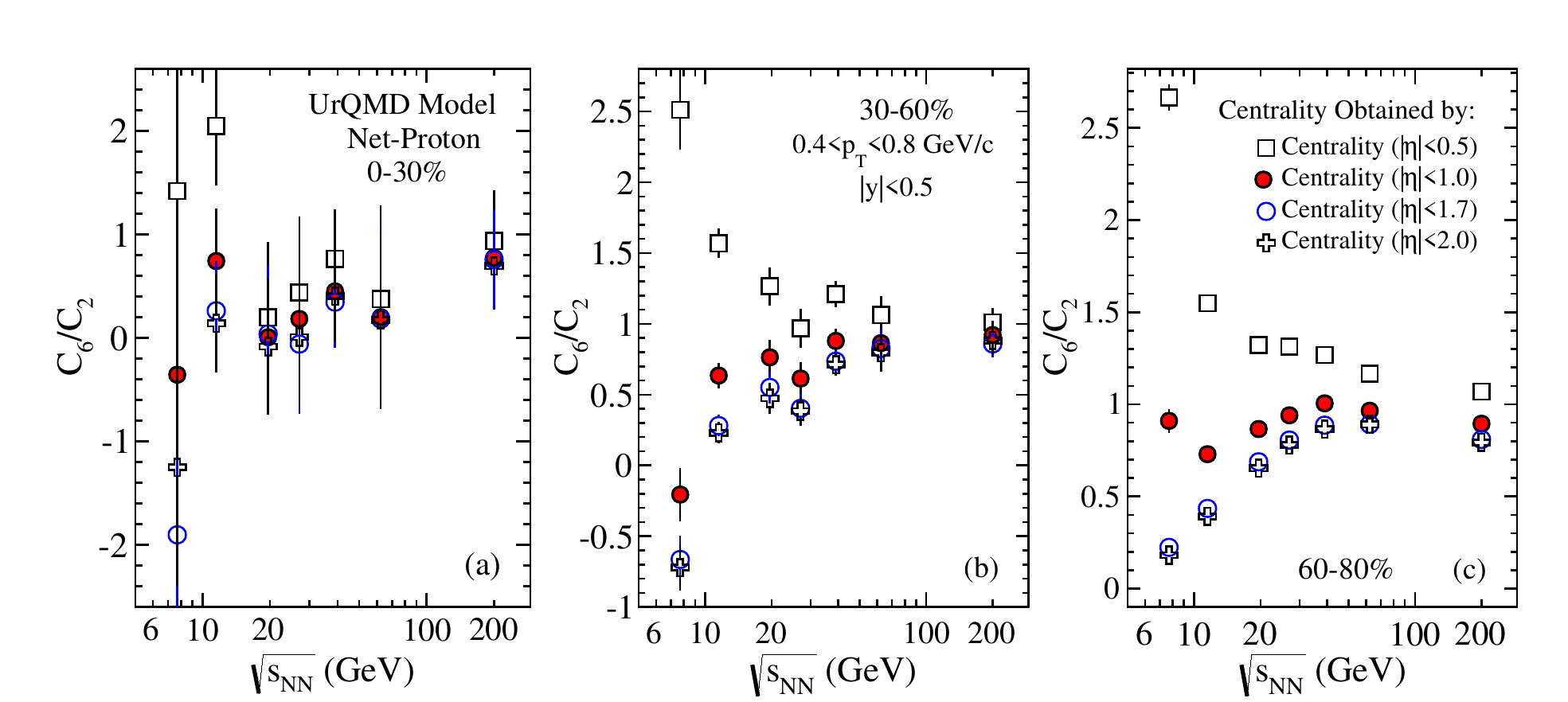}
\caption{\label{UrQMD-C6-resolution}  (Color online) Energy dependence of $C_6/C_2$ in central (0-30\%), mid-central (30-60\%), and peripheral ( 60-80\%) Au + Au collisions using the UrQMD model. The centralities are determined by the number of charged $\pi$ and $K$ within different $|\eta|$ ranges: $|\eta|<0.5, 1.0, 1.7$, and 2.0, respectively.}
\end{figure}

To study $C_6/C_2$ using the UrQMD model, the number of minimum bias events shown in Table~\ref{UrQMD-statistics} are generated. The number of events are comparable to those proposed for the future RHIC BES II~\cite{BESII-statistics}.
Furthermore, they are already larger than required by the CLT  based on the CWBC method in each $\delta1\%$ centrality bin.
Fig.~\ref{UrQMD-C6-resolution} shows the influence of  the centrality determinations on the measurement of  $C_{6}/C_{2}$ in central (0-30\%), mid-central (30-60\%) and peripheral (60-80\%)  Au + Au collisions  using the UrQMD model.
The kinematic intervals of  transverse momentum and rapidity are $0.4 < p_T < 0.8$ GeV/$c$ and $|y| < 0.5$,
which are the same as being used in published STAR measurement~\cite{STAR-proton}.
 With the larger acceptance, $0.4<p_T<2.0$ GeV/$c$, $C_6/C_2$ is not presented due to even larger uncertainties, although the statistics is larger than required by the  CLT  and the statistics are comparable to that expected for the RHIC BES II.
The centralities  are determined by the number of charged $\pi$ and $K$ produced in the final state with four typical $|\eta|$ ranges: $|\eta|<$ 0.5, 1.0, 1.7, and 2.0. $|\eta|<1.0$ is the current acceptance of the STAR detector, while $|\eta|<1.7$ is the acceptance for the upgraded STAR detector during RHIC BES II~\cite{BESII-statistics, BESII-acceptance}. $|\eta|<2.0$ is the acceptance which can detect most of the particles.

As shown in Fig.~\ref{UrQMD-C6-resolution}(a), for central collisions (0-30\%), $C_6/C_2$
obtained from a centrality determined within  $|\eta|<0.5$ is slightly larger than the other three cases.  The results are closer to each other with centralities determined by the  number of charged $\pi$ and $K$ within $|\eta|<$1.0, 1.7, and 2.0.

In mid-central and peripheral collisions, Fig.~\ref{UrQMD-C6-resolution}(b) and (c) show  that $C_6/C_2$ decreases as energy increases  with the centrality determined by the number of charged $\pi$ and $K$ within $|\eta|<0.5$. Most of the values are larger than unity.
As $|\eta|$ range for centrality definition becomes wider  (such as $|\eta|<1.0, 1.7$, and 2.0),
the trend of  the energy dependence of $C_6/C_2$ is significantly different from that obtained from $|\eta|<0.5$.
The strong dependence on the centrality determination in peripheral collisions is similar with  $\kappa\sigma^2$ \cite{Xiaofeng-JPG}.
 The significant difference indicates that the centrality resolution with $|\eta|<0.5$ is too poor to study $C_6/C_2$. On the other hand, it is also found that the values of $C_6/C_2$ with centrality determined by $|\eta|<1.0$ are still systematically larger than that corresponding to centralities defined by $|\eta|<1.7$ and 2.0.  The lower the energy, the larger the differences of measured values of $C_6/C_2$. Therefore, quantitative values of  $C_6/C_2$ in mid-central and peripheral collisions are not meaningful if  $|\eta|<1.0$ is used for the centrality determination.

 In conclusion, the centrality resolution with $|\eta|<0.5$ is too poor to study $C_6/C_2$. Currently, $|\eta|<1.0$ is the most acceptable acceptance for centrality determination with the STAR detector at RHIC.
  In this case, $C_6/C_2$ is consistent with that obtained from $|\eta|<1.7$ and 2.0 within statistical uncertainties in central collisions, although the results at mid-central and peripheral collisions are still slightly overestimated. In the upcoming RHIC BES II measurements, we can well the extract energy dependence of $C_6/C_2$ at all these three typical centralities.
In the UrQMD model, Fig.~\ref{UrQMD-C6-resolution} shows $C_6/C_2$ exhibits no energy dependence in central collisions within large uncertainties.

\section{ Discussions on signal observation at the upcoming RHIC BES II}
 The previous section shows no observation of a energy dependency of $C_6/C_2$  in central collisions using UrQMD model calculations.
Since the data generated with the UrQMD model are comparable to that expected at the upcoming RHIC BES II,  we believe that the
experimental uncertainties of the $ C_6/C_2$ measurement will not be small either.
 On the other hand, Lattice QCD  calculations and the PQM model  predict a strong negative signal of  $C_6/C_2$ near the  phase transition~\cite{QCD-9, PQM}. In this section, we want to study whether the predicted transition signal  could be observed at the upcoming RHIC BES II.

\begin{figure}[htmp]
\includegraphics[width=1.75in]{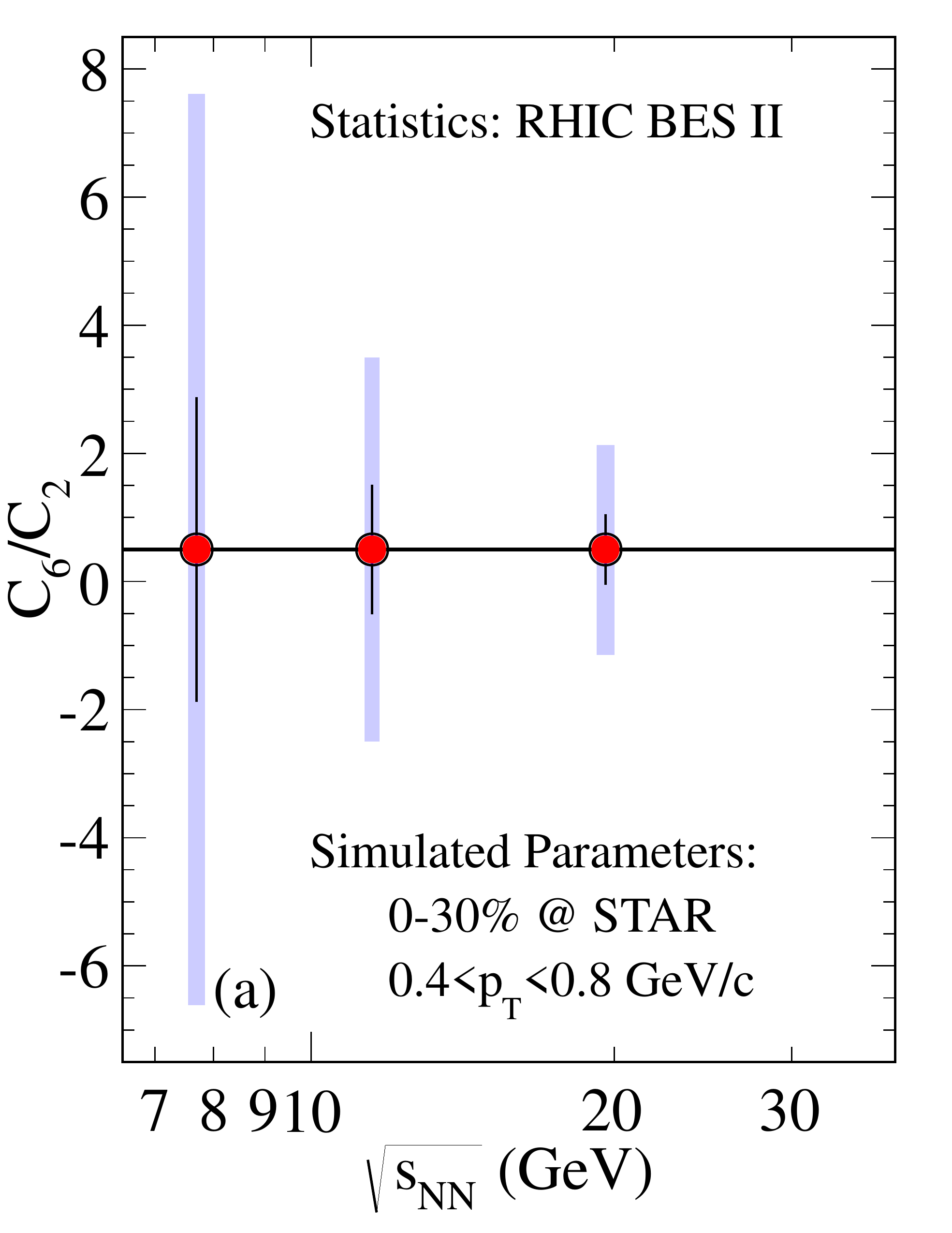}
\includegraphics[width=1.75in]{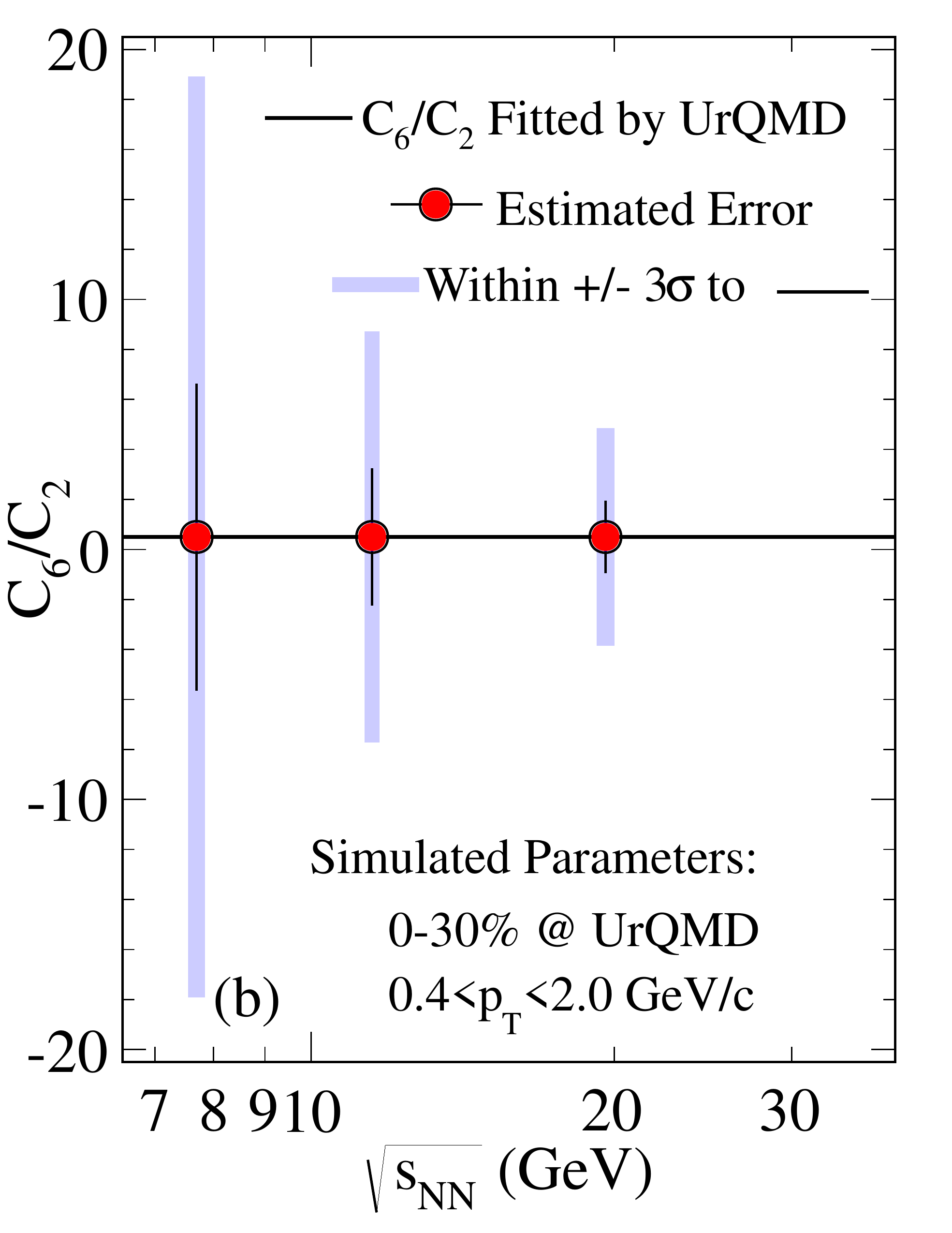}
\includegraphics[width=1.6in]{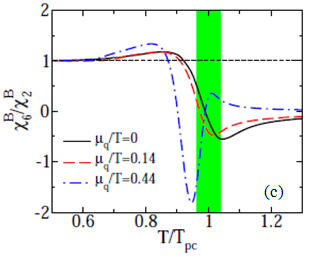}
\caption{\label{C6-3deviation}  (Color online)  The  blue areas in the left and middle panels show $C_6/C_2$ within 3$\sigma$ signal deviation to the black solid line with momentum range $0.4<p_{T}<0.8 $ GeV/$c$ and $0.4<p_{T}<2.0$ GeV/$c$, respectively.   The black solid line is value of $C_6/C_2$ fitted by a straight line using the UrQMD model.  The right panel is taken from Ref~\cite{PQM}. It shows
the temperature dependence of $\chi_6^B/\chi_2^B$   for various $\mu_q/T$  in the PQM model.
}
\end{figure}

The proposed numbers of events in 0-30\% central collisions for STAR during the RHIC BES II are about 30M, 80M, and 130M at $\sqrt{s_{NN}}$ = 7.7, 11.5, and 19.6 GeV, respectively~\cite{BESII-statistics}. Assuming  the experimental systematical uncertainties  are the same as the statistical uncertainties,
the total uncertainties of $C_6/C_2$ with $0.4<p_T<2.0$ GeV/$c$ and $0.4<p_{T}<0.8$ GeV/$c$ are shown in Fig.~\ref{C6-3deviation}(a) and (b), respectively.
The value of  the black straight solid lines is 0.5, which corresponds to the value of  $C_6/C_2$ fitted by a  straight line in 0-30\% central collisions using the UrQMD model with the  centrality  determined within $|\eta|<1.0$.   The expected $C_6/C_2$ within $\pm 3\sigma$ as deviation from 0.5 is shown by blue areas in Fig.~\ref{C6-3deviation}(a) and (b). It
is clearly seen  that only observing a negative $C_6/C_2$ is not sufficient even at the upcoming RHIC BES II. A conclusion can only be drawn if the value of $C_6/C_2$ is outside the $\pm 3\sigma$ band.

Ref~\cite{PQM} demonstrated $C_6/C_2$ as a function of  $T/T_{pc}$
in the PQM model at $\mu_q/T$ = 0, 0.14, and 0.44,  shown in Fig.~\ref{C6-3deviation}(c). The parameter $T_{pc}$ is pseudo-critical temperature corresponding to a peak in the chiral susceptibility in QCD
with physical light quark masses~\cite{PQM}.
Based on parameterized formulas of the energy dependence of the baryon chemical potential~\cite{Redilich-chemical},
  $\mu_q/T = $ 0.14 and 0.44 are corresponding to $\sqrt{s_{NN}} =$ 62.4 and 19.6 GeV, respectively.
 At $\sqrt{s_{NN}}=19.6$ GeV, the minimum value of  $C_6/C_2$  in the critical region, shown by green shaded area, is around -1.5.  If the energy dependence of $C_6/C_2$ can be observed, the difference of $C_6/C_2$ to 0.5, $\left|C_6/C_2 - 0.5\right|$, should be larger than 3$\sigma$,
where $\sigma$ is the corresponding uncertainty of $C_6/C_2$.

 In Fig.~\ref{C6-3deviation}(a), the  uncertainties at $\sqrt{s_{NN}}$ = 7.7, 11.5, and 19.6 GeV are 2.38, 1.01, and 0.55, respectively.
At $\sqrt{s_{NN}}=$ 19.6 GeV,  the difference of -1.5, which is the minimum value of $C_6/C_2$ in the critical region in the PQM model, to 0.5 is about 3.5$\sigma$.
Within the transverse momentum range $0.4<p_{T}<0.8$ GeV/$c$, one  can expect that the produced particles carry the information of a phase transition.
Consequently,
it seems to be possible to observe the  energy dependence of $C_6/C_2$ at $\sqrt{s_{NN}}=$19.6 GeV if the transition signal is really as strong as predicted by the PQM model.

The uncertainties at $\sqrt{s_{NN}} =  $ 7.7 and 11.5 GeV shown in Fig.~\ref{C6-3deviation}(a) are   significantly    larger than that at $\sqrt{s_{NN}}=19.6$ GeV.
Although Ref~\cite{PQM} did not present $C_6/C_2$ at these two energies,
it may still be a challenge  to study the energy dependence of $C_6/C_2$ at these two energies even if a strong transition signal exists.
On the other hand, it is better to use
$C_6/C_2$ obtained from Lattice QCD calculations to make an estimate  if the proposed number of events for RHICH BES II are sufficient. Currently, values of $C_6/C_2$ obtained from Lattice QCD calculations are with large uncertainties, although negative results are observed~\cite{QCD-9}.
Precise $C_6/C_2$ values based on QCD calculations with smaller uncertainties are urgent to be obtained before the upcoming RHIC BES II.

Fig.~\ref{C6-3deviation}(b) clearly shows that the transitional signal of $C_6/C_2$ in the PQM model can be totally washed out if $C_6/C_2$ is obtained within  $0.4<p_{T}<2.0$ GeV/$c$. Even at $\sqrt{s_{NN}}$ = 19.6 GeV, if $3\sigma$ of deviation from 0.5 is expected to be observed, the required statistics is about  600M events,  according to delta theorem calculations.

\section{Summary}
In summary, it is  interesting to study the energy dependence of $C_6/C_2$ from  RHIC BES to LHC energies.
In this paper, we study the non-critical background contributions and the required statistics for measuring $C_6/C_2$ with STAR at RHIC.
The CBWC method is suggested to reduce the initial size fluctuations on cumulant.
 Two major points to perform the CBWC method are that the selected centrality bin width should optimally reduce the initial size fluctuations and that the statistics in the corresponding bin  should satisfy the CLT.  With current available number of events,
 the CBWC method in each $1\%$ centrality bin is suggested to be applied for the $C_6/ C_2$ when $\sqrt{s_{NN}}\ge $19.6 GeV. At low energies, especially at $\sqrt{s_{NN}}$= 7.7 and 11.5 GeV,
there is no suitable centrality bin width to apply the CBWC method.
 With future proposed statistics for RHIC BES II, the $\delta1\%$ centrality bin width can be well applied to obtain $C_6/C_2$ of net-proton multiplicity distributions.

Based on the suggested  CBWC method in each $\delta1\%$ centrality bin,
the influence of different centrality determinations of $C_6/C_2$ are studied.
Currently, the centrality  determined within $|\eta|<1.0$ is the most acceptable choice to study $C_6/C_2$ with  STAR  at RHIC.
 At the upcoming RHIC BES II, we can  extract the energy dependence of $C_6/C_2$ based on centrality defined by number  of charged $K$ and $\pi$ produced in the final state within $|\eta|<1.7$.
With statistics comparable to the expected one of BES II, no signal of the energy dependence of $C_6/C_2$ is observed  in central collisions using the UrQMD model.

With the estimated uncertainties at the upcoming RHIC BES II, it is found that observing only a negative $C_6/C_2$ value is not sufficient.
It can be hidden within the experimental uncertainties.  If the transition signal is as strong as that in the PQM model,
 it is still challenging to observe it within  $0.4<p_T<2.0$ GeV/$c$.
However,  the deviation  is about $3.5\sigma$ within $0.4<p_{T}<0.8 $ GeV/$c$. Within this transverse momentum range, the produced particles are expected to carry information of a phase transition. Therefore, the transition signal of $C_6/C_2$ can hopefully be observed in the measurement.

\section{Acknowledgements}
This work is supported in part by the Major State Basic Research Development Program of China under Grant No. 2014CB845402, the NSFC of China under Grants  No. 11405088, No. 11221504, No. 11005046,  and the Ministry of Education of China with Project No. 20120144110001. The second author thanks Chinese Scholarship Council for financial support No. 201506775038.


\begin{thebibliography}{99}
\bibitem{QCD-1} M. A. Stephanov, K. Rajagopal, and E. V. Shuryak, Phys. Rev. Lett. {\bf 81}, 4816 (1998); Phys. Rev. D {\bf 60}, 114028 (1999).
\bibitem{QCD-2} Z. Fodor and S. D. Katz, J. High Energy Phys. {\bf 04}, 050 (2004).
\bibitem{QCD-3} S. Ejiri, Phys. Rev. D {\bf 78}, 074507 (2008).
\bibitem{QCD-4} M. Cheng {\itshape et al}., Phys. Rev. D {\bf 79}, 074505 (2009).
\bibitem{QCD-5} M. A. Stephanov, Phys. Rev. Lett. {102}, 032301 (2009); Phys. Rev. Lett. {\bf 107}, 052301 (2011).
\bibitem{QCD-6} F. Karsch, K. Redlich. Phys. Lett. B {\bf 695}, 136 (2011).
 \bibitem{QCD-7} S. Gupta {\itshape et al}., Science {\bf 332}, 1525 (2011).
\bibitem{QCD-8} R. V. Gavai and S. Gupta, Phys. Lett. B {\bf 696}, 459 (2011).
\bibitem{slowing-1} B. Berdnikov and K. Rajagopal, Phys.Rev. D {\bf61}, 105017 (2000).
\bibitem{slowing-2} S. Mukherjee, R. Venugopalan, and Y. Yin,  arXiv: 1506.00645[hep-ph].
\bibitem{STAR-proton} L. Adamczyk {\itshape et al}., (STAR Collaboration), Phys. Rev. Lett {\bf 112}, 032302 (2014).
\bibitem{STAR-charge} L. Adamczyk {\itshape et al}., (STAR Collaboration), Phys. Rev. Lett {\bf 113}, 092301 (2014).
\bibitem{phenix-charge} A. Adare {\itshape et al}., (PHENIX Collaboration), Phys. Rev. C {\bf 93}, 011901(R) (2016).
\bibitem{QCD-9} C. Schmidt, Prog. Theor. Phys. Suppl. {\bf 186}, 563 (2010).
\bibitem{PQM} B. Friman {\itshape et al}., Eur. Phys. J. C {\bf71},1694 (2011).
\bibitem{panxue} X. Pan {\itshape et al}.,  Nucl. Phys. A {\bf913}  206 (2013).
\bibitem{o4} K. Morita, K. Redlich, arXiv: 1409.8001[hep-ph].
\bibitem{star} J. Adams {\itshape et al}., (STAR Collaboration), Nucl. Phys. A {\bf 757}, 102 (2005).
\bibitem{lhc} K. Aamodt {\itshape et al}., (ALICE Collaboration), JINST {\bf 3} S08002 (2008).
\bibitem{conservation} A. Bzdak, V. Koch and V. Skokov. Phys. Rev. C {\bf 87}, 014901 (2013).
\bibitem{efficiency-1} A. Bzdak and V. Koch. Phys. Rev. C {\bf 86}, 044904 (2012).
\bibitem{efficiency-2} A. Bzdak and V. Koch. Phys. Rev. C {\bf 91}, 027901 (2015).
\bibitem{initial-1} Henning Heiselberg. Physics Reports {\bf 351}, 161 (2001).
\bibitem{Xiaofeng-JPG}X. Luo {\itshape et al}.,  J. Phys. G {\bf 40}, 105104 (2013).
\bibitem{chenlz-JPG} L. Chen, Z. Li and Y. Wu, J. Phys. G {\bf 41}, 105107 (2014).
\bibitem{Nihar-charge} N. R. Sahoo, S. De, and T. K. Nayak, Phys. Rev. C {\bf 87}, 044906 (2013).
\bibitem{PHENIX-c} S. S. Adler {\itshape et al}., {PHENIX Collaboration}, Phys. Rev. C, {\bf 71}, 034908 (2005).
\bibitem{CBWC-STAR} X. Luo (For the STAR Collaboration), J. Phys. : Conf. Ser. {\bf316}, 012003 (2011).
\bibitem{CLT} John A. Rice, Mathematical Statistics and Data
Analysis (Third ed.), Duxbury Press, ISBN 0-534-39942-8.
\bibitem{Skellam-1} Skellam J G, Journal of the Royal Statistical Society,
{\bf 109}, 296 (1946).
\bibitem{Skellam-2} P. Braun-Munzinger et al., Phys. Rev. C {\bf84}, 064911
(2011).
\bibitem{Skellam-3} X. Pan {\itshape et al}., Phys. Rev. C {\bf 89}, 014904 (2014).
\bibitem{xiaofeng-delta}  X. Luo,  J. Phys. G {\bf 39}, 025008 (2012).
\bibitem{UrQMD} M. Bleicher{\itshape et al}., J. Phys. G {\bf 25}, 1859 (1999).
\bibitem{BESII-statistics}D. Cebra talk at Critical Point and Onset of Deconfinement,  University of Bielefeld, Germany, November 17-21, 2014. http://www2.physik.uni-bielefeld.de/cpod2014.html.
\bibitem{BESII-STAR}STAR Collaboration, ��Studying the Phase Diagram of QCD Matter at RHIC,�� STAR Notes
SN0598. https://drupal.star.bnl.gov/STAR/starnotes/public/sn0598 (2014).
\bibitem{BESII-acceptance} Y. Wang (for the STAR Collaboration), Journal of Physics: Conference Series {\bf535}, 012022 (2014).
\bibitem{Redilich-chemical} J. Cleymans {\itshape et al}., Phys. Rev. C {\bf73}, 034905 (2006).

\end{thebibliography}
\end{document}